\documentclass{elsart}
\usepackage{graphicx,harvard,amssymb}

\makeatletter
\def\elsartstyle{%
        \def\normalsize{\@setfontsize\normalsize\@xiipt{14.5}}
        \def\small{\@setfontsize\small\@xipt{13.6}}
        \let\footnotesize=\small
        \def\large{\@setfontsize\large\@xivpt{18}}
        \def\Large{\@setfontsize\Large\@xviipt{22}}
        \skip\@mpfootins = 18\p@ \@plus 2\p@
        \normalsize
}
\makeatother

\def\url#1{{\ttfamily\def\/{/\discretionary{}{}{}}#1}}

\def\ni{\noindent}

\def\ea{et al. \,}
\def\eg{{\it e.g. \,}}
\def\be{\begin{equation}}
\def\ee{\end{equation}}

\def\ls{\mathrel{\raise0.35ex\hbox{$\scriptstyle <$}\kern-0.6em
    \lower0.40ex\hbox{{$\scriptstyle \sim$}}}}
\def\gs{\mathrel{\raise0.35ex\hbox{$\scriptstyle >$}\kern-0.6em 
    \lower0.40ex\hbox{{$\scriptstyle \sim$}}}}

\begin{document}

\begin{frontmatter}
\title{Dark Matter Profiles in Clusters of Galaxies: a Phenomenological 
Approach}

\author{Yinon Arieli}
\address{School of Physics and Astronomy, Tel Aviv University, Tel 
Aviv, 69978, Israel}
\author{Yoel Rephaeli}
\address{School of Physics and Astronomy, Tel Aviv University, Tel 
Aviv, 69978, Israel, \\ Center for Astrophysics and Space 
Sciences, University of California, San Diego, La Jolla, 
CA\,92093-0424}

\vspace{0.7cm}
\noindent \small{{\it Accepted for publication in New Astronomy}}
\thanks[email]{E-mail: yinonar@post.tau.ac.il}
\thanks[email]{E-mail: yoelr@noga.tau.ac.il}

\begin{abstract}

There are some basic differences between the observed properties 
of galaxies and clusters and the predictions from current 
hydrodynamical simulations. These are particularly pronounced 
in the central regions of galaxies and clusters. The popular NFW 
(Navarro, Frenk, \& White) profile, for example, predicts a density 
cusp at the center, a behavior that (unsurprisingly) has not been 
observed. While it is not fully clear what are the reasons for this 
discrepancy, it perhaps reflects (at least partly) insufficient 
spatial resolution of the simulations. In this paper we explore a purely 
phenomenological approach to determine dark matter density profiles 
that are more consistent with observational results. Specifically, 
we deduce the gas density distribution from measured X-ray brightness 
profiles, and substitute it in the hydrostatic equilibrium equation 
in order to derive the form of dark matter profiles. Given some 
basic theoretical requirements from a dark matter profile, we then 
consider a number of simple profiles that have the desired asymptotic 
form. We conclude that a dark matter profile of the form 
$\rho=\rho_0\left(1+r/r_a \right)^{-3}$ is most consistent with
current observational results. 
\end{abstract}

\begin{keyword}
dark matter profiles - Galaxies: clusters: general - X-rays: galaxies: clusters
\PACS 98.65.Cw ; 95.35.$+d$ ; 95.85Nv
\end{keyword}
\end{frontmatter}

\section{Introduction}

Mass density profiles of galaxies and clusters of galaxies
play a central role in the study of the intrinsic properties 
of these systems as well as in models of their formation, 
evolution, and their use as probes of the mass density of the 
universe. The formation of cold dark matter (CDM) halos has been 
studied extensively over the years. Considerable theoretical work 
has been done to describe the shape of DM profiles, mostly by 
numerical simulations (\eg, Suto 2002). Early attempts were 
severely limited in predicting the profile in the central region 
due to insufficient spatial resolution. Recent improvements in 
numerical techniques led to the attainment of high central 
resolution ($\sim 10$ kpc) in the dynamical simulation of DM 
profiles. However, a realistic description of the distribution 
of DM and gas in the central cluster regions necessitates a coupled 
dynamical and hydrodynamical simulation that can follow the evolution 
of DM and gas under their mutual interactions. In particular, heating 
and cooling processes in the gas can occur on timescales which are 
shorter than the Hubble time with possible ramifications also for 
the DM profile.

Based on large, N-body simulations of the evolution of DM configurations, 
Navarro, Frenk \& White (hereafter NFW, 1995, 1997) showed that CDM 
density profiles are independent of the halo mass, and can be accurately 
fit over a large range of sizes by a simple algebraic form which is said 
to be universal (but see Jing \& Suto 2000). The proposed NFW profile has 
a cusp-like $r^{-1}$ behavior close to the center, and an asymptotic 
$r^{-3}$ falloff at large $r$. It has been argued that DM profiles may even  
exhibit a steeper inner cusp; for example, Moore et al. (1999) claimed that 
high resolution simulations indicate that the central profile is 
$\propto r^{-1.5}$.

An uninterrupted steep rise of the density towards the center is 
clearly unphysical, and such a characterization could be due to 
insufficient level of spatial resolution in the numerical 
simulations, other important factors in the simulations (such as 
the particular choice of initial conditions, e.g., Bartschiger \& 
Labini 2001), or a result of physical limitations in the description 
of the cluster density over small (O[10 kpc]), typically {\it 
galactic} scales. That the latter are perhaps the more likely reasons 
is indicated by even a steeper inner cusp that is deduced in recent 
(Governato et al. 2001) higher resolution simulations. Even the
improved hydrodynamical simulations do not include all the relevant 
physical processes that could affect the nature of the deduced mass 
profiles. For example, it is clear that the properties of intracluster
(IC) gas, whose fractional contribution to the total mass is $\sim
10\%$, must play some role in the determination of the density profile.

Recent {\it Chandra} observations of the central regions of a few nearby
clusters have not (yet) provided unequivocal evidence on the shape of 
the central DM profile. Strong evidence was found for a flat profile
in the central region of Abell 1795 (Ettori \ea 2002), a behavior
similar to the trend seen in observations in some galaxies (mainly low
surface brightness and dwarf galaxies). But there is also evidence for 
a central cusp in the clusters Abell 2029 (Lewis et al 2002), Hydra A
(David \ea 2001), and EMSS 1358 (Arabadjis \ea 2002). However, evidence 
for the latter is weak, given the low quality of the fits (generally large 
$\chi^2/dof$), central CD galaxy (in Abell 2029), low central resolution 
(in EMSS 1358), and relatively high value of the concentration parameter 
with respect to expectations from numerical simulations (in Hydra A). 
Clearly, many more {\it Chandra} and XMM observations of the central 
regions of clusters are needed in order to discern a clear trend in the 
shape of DM profiles there.

Over the last few years the NFW profile has been adopted in
calculations of the structure and evolution of CDM halos, this
in spite of its unappealing central cusp. There is a clear need for 
further exploration of cluster mass profiles with the aim of either 
modifying the NFW profile, or finding an alternative profile that is
more consistent with observations. The inconsistency between observational 
results and predictions from simulations provides considerable 
motivation for a more {\it phenomenological} approach that is based 
on dynamical deductions from the observed properties of IC gas, an 
approach that is adopted in this paper. The gas density and temperature 
profiles can be determined from current high quality X-ray measurements; 
these profiles can then be used to probe the total density distribution.

We begin with a short review of IC gas density profiles,and their use 
to probe the DM profile based on the hydrostatic equilibrium (HE) equation. 
In section 3 we describe the limitations of the NFW profile for the IC DM 
density. Alternative DM profiles are discussed in section 4; we
consider the requirements from a DM profile, and 
attempt a solution to the divergence problem of the NFW profile in 
section 4.1. This consists of a slight but physically important modification 
of the NFW profile which results in a finite density at the center. A 
theoretical discussion of new phenomenological DM profiles is given in 
section 4.2. Next (section 5), we confront several different DM
profiles with observational data, primarily a sample of ROSAT 
measurements of 24 nearby and moderately distant 
clusters at redshifts $z\leq 0.2$, 
and draw some conclusions on the form of the most viable profile. In 
Section 6 we summarize and briefly discuss a few other aspects of the 
subject matter.

\section{Spatial Distribution of IC Gas}

Spectral measurements of thermal bremsstrahlung X-ray emission from
IC gas provide an integrated
measure of the emissivity-weighted temperature, while the gas
density profile is deduced from measurements of the surface
brightness (SB) distribution across the cluster.
A starting point in a theoretical
description of the gas in a relaxed cluster is the attainment of
HE, and although this includes aspehrical
configurations, the assumption of spherical symmetry is still
reasonable for at least a subset of rich and regular clusters.
The expected availability of uniform datasets of measurements of many
clusters with the XMM and {\it Chandra} satellites motivates
a more realistic modeling of the gas thermal and spatial distributions
than afforded by an isothermal $\beta$ model. An example for a more 
general description is a polytropic equation of state 
$P \propto \rho_g^{\gamma}$ relating the (thermal, assumed dominant) 
pressure and ({\it gas}) density, with the index $\gamma$ as a free 
parameter. The HE equation is then
\begin{equation}
\frac{k T_{0}\gamma}{\mu m_{p}
 \rho_{g0}^{\gamma-1}}\rho_g^{\gamma-1}\frac{d ln\rho_g}{dr}=-\frac{G
  M(r)}{r^2}\,,
\end{equation}
where k, $\mu$ and $m_{p}$ are the Boltzmann constant, the mean 
molecular weight, and the proton mass, respectively; $M(r)$ is the 
total cluster mass interior to $r$ -- a sum of the masses of DM, gas 
and galaxies.

The gas density is usually represented by an analytic (King) $\beta$ 
profile
\begin{equation}
\rho_g(r)=\frac{\rho_{g0}}{\left[1+y^2\right]^{3 \beta/2}}\,,
\end{equation}
with $y=r/r_c$; $r_c$ is the gas core radius. In the case of 
isothermal gas, 
$\gamma = 1$, the (sky) projected X-ray SB profile that corresponds to 
this density is of the form
\begin{equation}
S_x^{\beta}(R)=S_0\left(1+\frac{R^2}{r_c^2}\right)^{-3\beta+1/2}\,,
\end{equation}
where $S_0$ is the central SB, $R$ denotes the projected distance 
from the cluster center, and $\beta$ is a fit parameter.
The X-ray deduced gas parameters are used in the HE equation to 
determine $M(r)$. In the simplest treatment the gas contribution to 
the gravitational field can be ignored, to first approximation, since 
the gas mass fraction is $\leq 10\%$.

\section{The NFW Profile}

The NFW DM profile is 
\begin{equation}
\rho^{NFW}(x)=\frac{\rho_{0}}{x \left( 1+x
  \right)^2}\,,
\end{equation}
where $x = r/r_{s}$; $r_{s}$ is a scaling radial parameter. Both 
$r_{s}$ and the central density $\rho_0$ are related to the 
cosmological parameters (NFW 1997). 
While it is generally deduced from N-body simulations that the DM 
density is asymptotically $\propto r^{-3}$, the behavior of the 
NFW profile in the inner core is problematic: There is no 
clear observational evidence for such density cusps in clusters. Specifically, the deduced mass 
distribution in clusters disagrees with 
that deduced from the NFW profile, and the X-ray SB 
calculated from this profile has a much smaller core radius than 
deduced from observations (Suto et al. 1998). Of course, the 
rise of density can be truncated below a certain inner radius and 
replaced by a constant value; this, however, is too arbitrary. The
discrepancy is even larger in cD clusters.

\begin{figure}
\begin{center}
\includegraphics*[width=6cm]{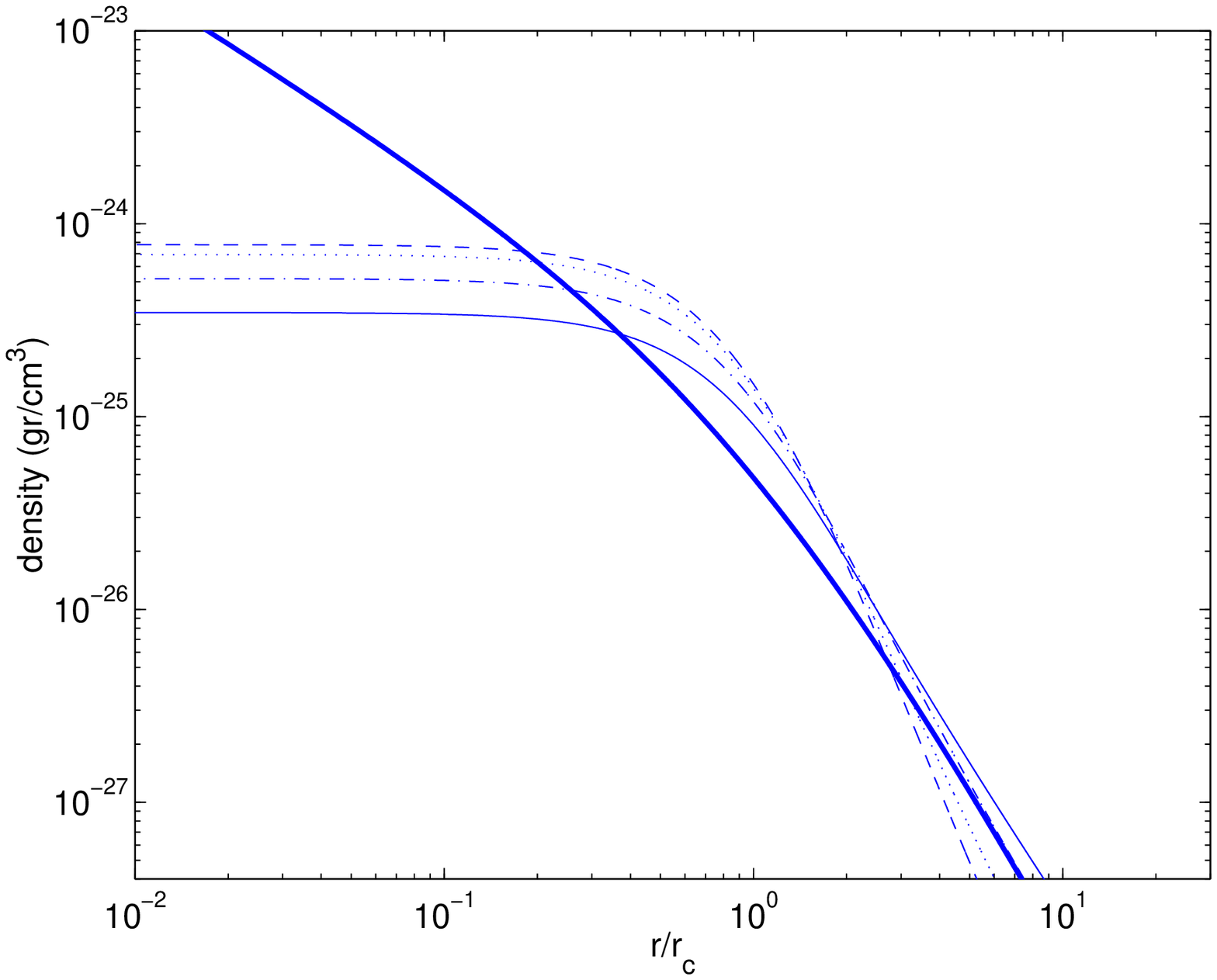}
\includegraphics*[width=6cm]{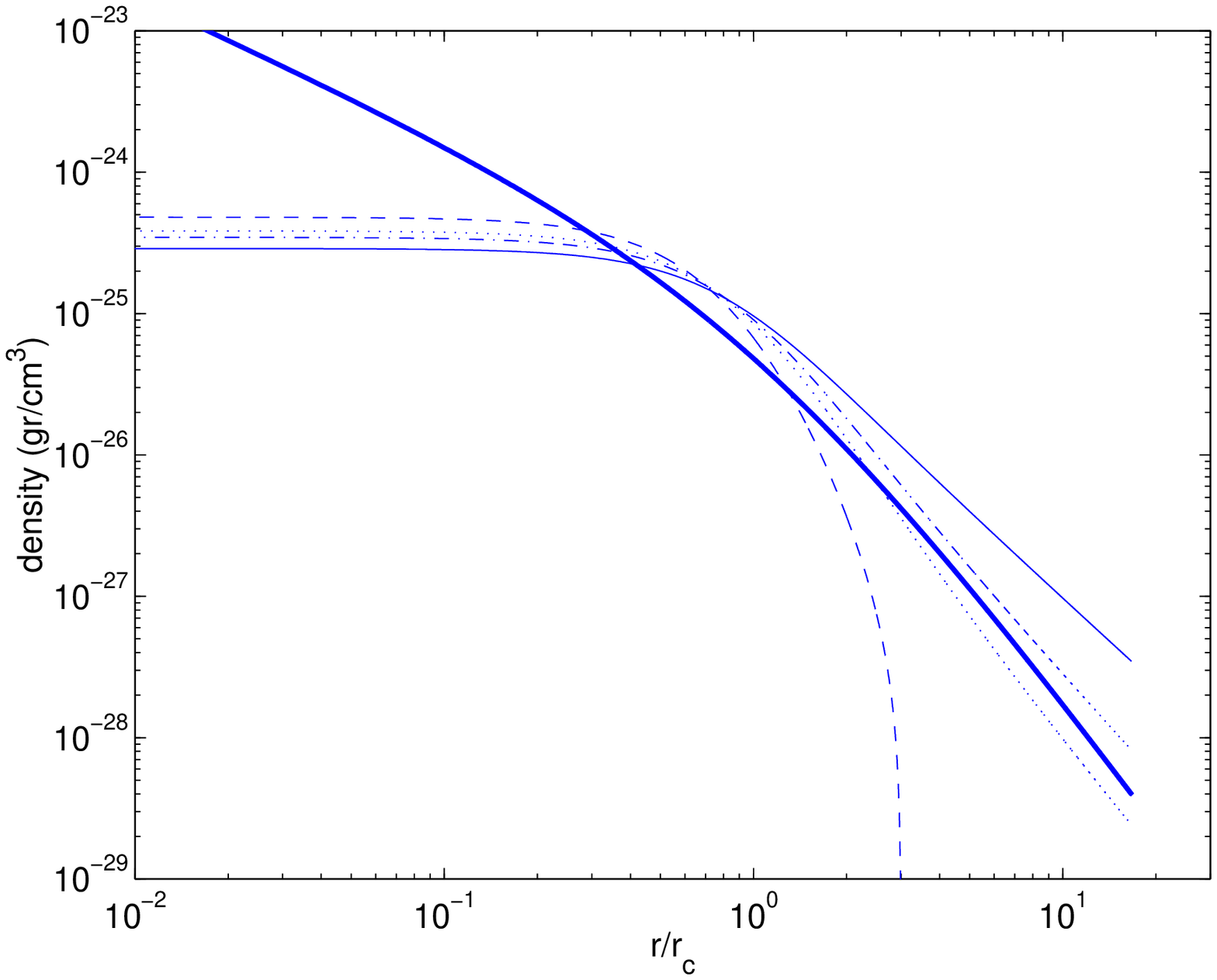}
\caption{DM density distribution obtained from the solution
  of the HE equation in the isothermal (left panel) and the more 
  general polytropic (right panel) case are shown for various values 
  of $\beta$ and $\gamma$. Solid, dotted-dashed, dotted
  and dashed lines correspond to the results of the calculations for
  $\beta$=2/3, 1, 4/3, and 3/2 in the isothermal case, and $\gamma$=1,
  1.2, 4/3 and 5/3 in the polytropic case with $\beta=2/3$, respectively. 
  The thick solid line in each of the panels depicts the NFW profile.}  
\end{center}
\end{figure}

\subsection{DM profile from polytropic gas}

We first show that there are appreciable differences between the DM 
profile deduced from an isothermal $\beta$ gas and the NFW profile. 
The spherically symmetric HE equation -- in the limit when the gas 
contribution to the gravitational field can be ignored -- yields in 
this case
\begin{equation}
\rho^{poly}(y)=-\tilde{A} \frac{\left[\left( 1+y^2\right)^{-3
      \beta/2}\right]^{\gamma-1}\left\{\left[ -3
      \beta\left(\gamma-1\right)+1\right] y^2 +3\right\}}{\left(1+y^2
  \right) ^2}\,,
\end{equation}
where $y = r/r_c$ and
\begin{equation}
\tilde{A} = \frac{3 k \beta T_{0}}{4 \pi G \mu m_p r_c^2}\,.
\end{equation}

In Figure 1 we show the resulting DM profiles and the NFW profile 
for typical values of the parameters; the left panel is for 
isothermal gas, while the right panel shows the results for 
$\gamma=$ 1.2, 4/3 and 5/3. There is a noticeable difference 
between these profiles and the NFW profile, especially at $r\leq 
r_c$. The calculated profiles converge to a constant central 
density, while the NFW profile diverges in this region. In the 
outer region the falloff of the NFW profile is more moderate than 
those of the other profiles.

Next we calculate the gas profile deduced from the NFW model; 
substituting the NFW profile (4) into the isothermal HE equation we 
have (as was obtained already by Suto, Sasaki \& Makino 1998)  
\begin{equation}
\rho_{g}(r)=\rho_{g0}  exp[-Bf(r/r_{s})]\,,
\end{equation}
where
\begin{eqnarray}
f(x)=1-\frac{1}{x}ln(1+x)\,,
\end{eqnarray}
for the NFW profile, and B is the dimensionless parameter
\begin{eqnarray}
B\equiv\frac{4\pi G \mu m_{p} \rho_{0} r_{s}^2}{k T_{g0}}\,.
\end{eqnarray}

In Figure 2 we plot the gas profiles deduced from the NFW profile 
for three values of B that correspond to a wide temperature range. 
Also shown are best fits of each of the three curves to a $\beta$ 
profile. The right panel shows the ratio between each 
profile and its best-fit $\beta$ profile.
\begin{figure}
\begin{center}
\includegraphics*[width=6cm]{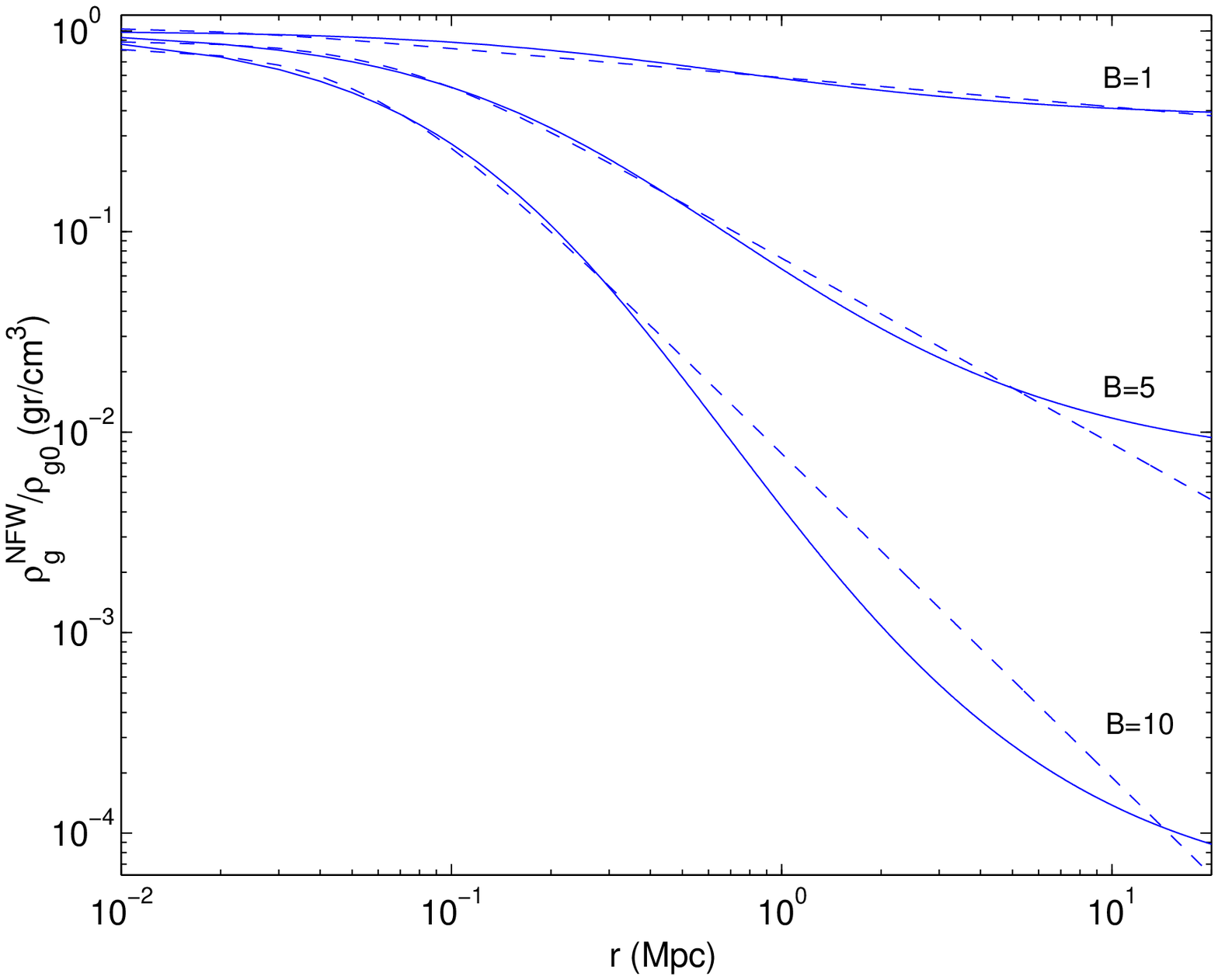}
\includegraphics*[width=6cm]{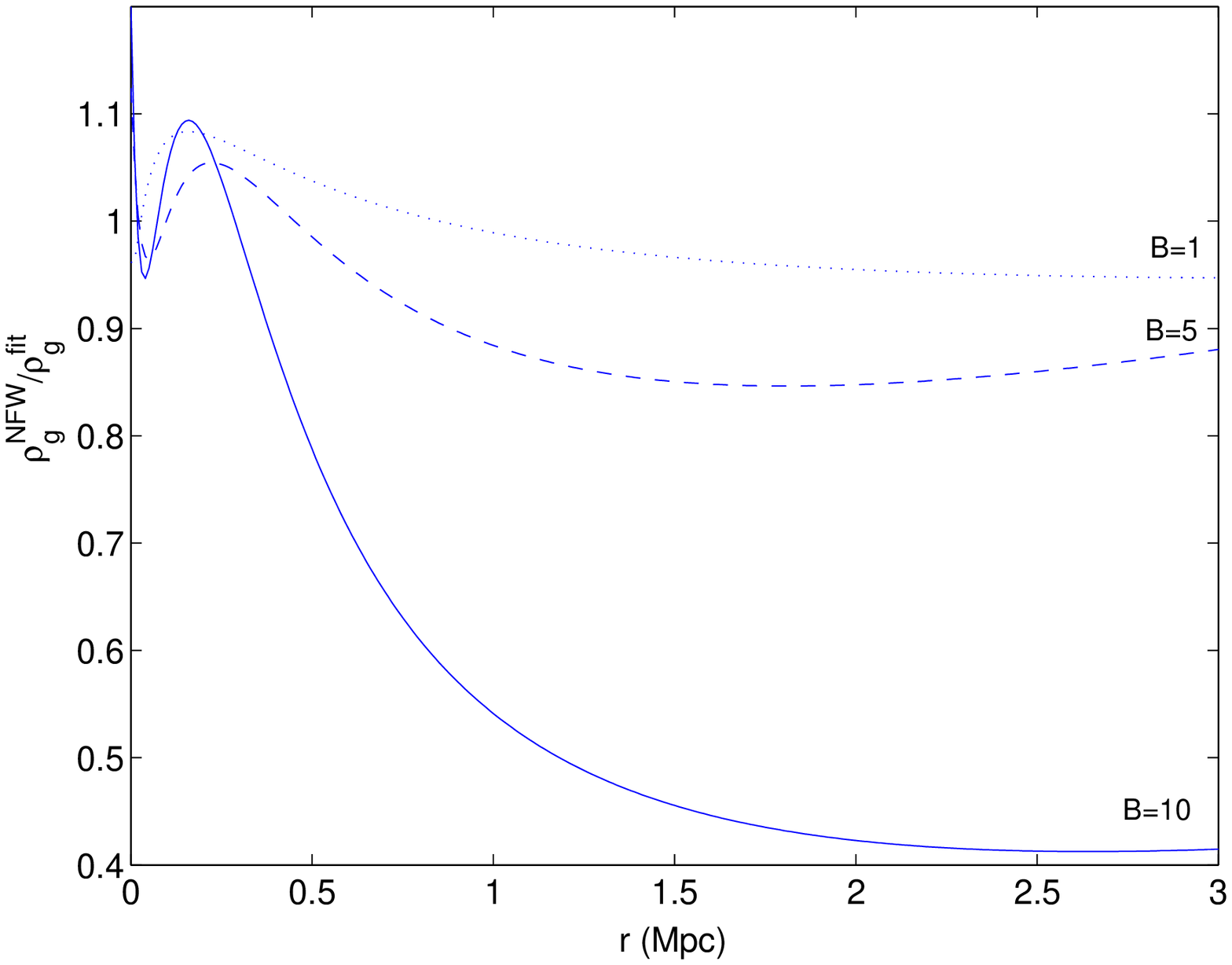}
\caption{The gas profiles deduced from substituting the NFW profile
  into the HE equation for different values of B (B=1, 5 and 10) are 
  plotted as solid lines in the left panel. 
  A dashed line next to each solid line shows the best fit to a 
  $\beta$ profile.
  The ratio between each profile and its best ($\beta$) fit is shown 
  in the right panel.}
\end{center}
\end{figure}
As can be seen in these figures the gas distribution deduced from
the NFW profile is quite different from the $\beta$ profile. Only in 
the extreme and highly unlikely case of small B $(B\approx 1)$ the 
two profiles are similar, with differences smaller than 10\%. 

Differences between the gas distribution deduced from the NFW 
profile and the $\beta$ profile can also be demonstrated by their 
related X-ray SB. Fitting this predicted SB profile to the observed 
profile, which is known to have a $\beta$ model shape, results in 
the dependence displayed in Figure 3.

\begin{figure}
\begin{center}
\includegraphics*[width=8cm]{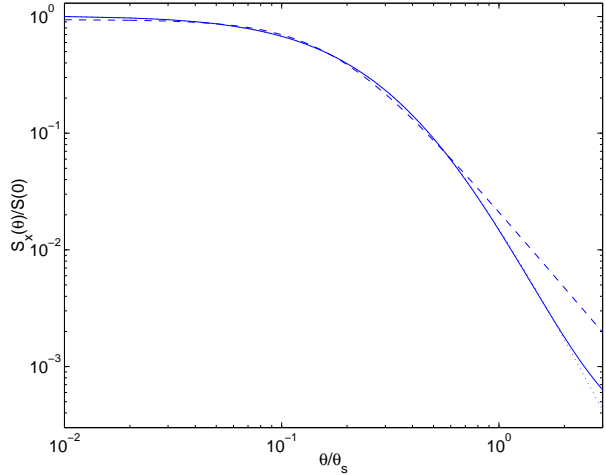}
\caption{The theoretical SB profile expected from substituting
  the NFW profile in the HE equation is plotted as solid line. The
  dashed and dotted lines are the best fit to a $\beta$ surface
  brightness profile and a general $\beta$ profile, respectively.}
\end{center}
\end{figure}
Clearly, the SB predicted from the NFW profile does not provide a 
reasonably good fit to the $\beta$-model SB function. Rather, we 
find that the former function can be well 
fit by a generalized $\beta$ distribution of the form
\begin{equation}
S_x^{new}(\theta/\theta_c)=\frac{S^{new}(0)}{\left[1+\left(\theta/\theta_c^{new}\right)^\xi
  \right]^{\tilde{\beta}}}\,,
\end{equation}
where $\xi \neq 2$, and $\tilde{\beta}\neq 3\beta-1/2$, i.e. values
different than those of the $\beta$ model 
(see dotted line in Figure 3).
These results are consistent with those of Suto et al. (1998). Thus, it 
is clear that the general SB function $S_x^{new}$ gives a better fit to 
$S_x^{NFW}$ than does $S_x^{\beta}$.

We conclude that 
the NFW profile seems to be inconsistent with X-ray 
observations of clusters. The main physical reason for the above 
differences is the 
excessively high DM concentration that is predicted by the NFW model
in the central cluster region.

\subsection{The NFW profile and polytropic gas}

When the NFW profile is substituted in the more general polytropic 
HE equation (1), the resulting gas density is 
\begin{equation}
\rho_g(r)=\rho_{g0}\left[1-\frac{B(\gamma-1)}{\gamma}f(r/r_s)
\right]^{\frac{1}{\gamma-1}}\,.
\end{equation}
This function assumes negative values beyond a critical radius and 
is therefore physically unacceptable at larger radii.

For a cluster with a maximal radius $r_{max}$, the requirement 
that the profile is non-negative constrains the value of the polytropic 
index to be 
\begin{equation}
\gamma \leq \frac{Bf(x_{max})}{Bf(x_{max})-1}\,,
\end{equation}
where we took $x=x_{max}$ since $f(x)$ is a monotonically increasing
function for the NFW profile. In table 1 we list the upper limit on the 
polytropic index for various values of $r_{max}$ and $B$, with a typical 
value of the scale radius, $r_s=0.2\,Mpc$. The results weakly depend on 
the value of the scale radius; for larger values of $r_s$, \eg, 
$0.5$ $Mpc$, the value of $\gamma_{max}$ increases only by $\sim 
5\%$. 
It would seem from Table 1 that in this model the polytropic index 
is limited to a relatively narrow range of values. 

\begin{table}
\begin{center}
\caption{Upper limit on $\gamma$ from the requirement that the
 gas profile is positive at $r \leq r_{max}$.}
\begin{tabular}{|c|c|c|c|c|c|c|c|c|c|} \hline\hline
$\gamma_{max}$ &  1.07& 1.16& 1.38& 1.07& 1.15& 1.35& 1.06& 1.13&
1.32  \\\cline{1-10}
$r_{max}$ (Mpc)    & 1.5& 1.5& 1.5& 2& 2& 2& 3& 3& 3 \\\cline{1-10}
B&       20& 10& 5&  20& 10& 5&  20& 10& 5  \\\cline{1-10}
\end{tabular}
\end{center}
\end{table}

\section{Alternative DM Profiles}

Having discussed the possibly 
problematic features of the NFW model, we now want to find alternative 
DM profiles that are more physically viable and are consistent with 
X-ray SB measurements. To do so we first 
specify the properties desired of a DM profile, and then consider 
a slightly modified NFW profile, and -- more generally -- 
the simplest functional forms that satisfy the requirements from a DM 
profile. We require that a DM profile has the following properties:

\ni
$1.$ Finite, positive-definite at all $r$.

\ni
$2.$ Asymptotic $r^{-3}$ behavior at large $r$.

\ni
$3.$ An associated gas profile (from HE equation) that has 
the form of a $\beta$-profile with a value of $\beta$ which is 
consistent with X-ray SB measurements.

The first property is an obvious physical requirement; the second 
is based on results of many N-body simulations, and the third is 
based on fits to observed SB profiles of many clusters. 
In assessing the viability of a DM profile we will also consider 
whether the central mass density is high enough for observable 
effects of gravitational lensing. In some clusters the measurements 
of either giant arcs, strong, or weak gravitational lensing imply 
that the DM central density needs to be sufficiently high to 
produce these lensing effects. Specifically, the central surface 
density has to be typically higher than the critical value of 
$\Sigma_0 \sim 0.5\, gr/cm^2$ (with $H_0=50\; km\; s^{-1}\; 
Mpc^{-1}$ and $\Omega=1$) in order that multiple lensed images are 
produced (Subramanian \& Cowling 1986). 

\subsection{A Modified NFW profile}

An operational approach to the divergence problem of the NFW 
profile is to replace it in the inner region, $r \leq r_b$, with a 
non-divergent form. 
From figures 1 \& 2, it is clear that in the 
central region of the cluster the NFW profile is much steeper than 
either $\rho^{iso}$ or $\rho^{poly}$, intersecting these curves 
inside the cluster core. Therefore, 
in order to 
remove the divergence of the NFW profile, in this inner region we 
replace the NFW profile with the functional form obtained as a 
solution --  eq. (5) -- to the HE equation when taking a 
$\beta$-profile for the gas. This form can then be tailored to the 
NFW profile outside the inner region, namely
\begin{eqnarray}
 \rho^{new-poly}(r)=\left\{ \begin{array}{ll} 
      \tilde{A} \frac{\left\{\left[ 1+(r/r_c)^2\right]^{-3
          \beta/2}\right\}^{\gamma-1}\left\{\left[-3
    \beta\left(\gamma-1\right)+1\right] (r/r_c)^2 +3\right\}}{\left[ 
    1+(r/r_c)^2 \right] ^2}
 & \mbox{$ r\leq r_{b}$} \\ \\
                     \frac{\rho_{0}}{(r/r_s) \left( 1+r/r_s
  \right)^2} & \mbox{$r > r_{b}$}
                  \end{array}
           \right.\,, 
\end{eqnarray}
for the (general) polytropic case.

The new modified profile is not smooth at $r = r_b$, but it is 
continuous at this point, and at larger radii it falls off 
asymptotically as $r^{-3}$. If we were to fit 
it by the functions of the kind of $\rho^{iso}$ and $\rho^{poly}$, the fit 
would be excellent in the central region, but the quality of the fit 
would deteriorate outside this region. Thus, the overall discrepancy 
between this modified DM profile and the profile deduced from the 
measured gas density would still remain. 
(Note that 
the DM profile can be similarly changed in clusters with a giant cD 
galaxy whose DM density distribution is generally different from the NFW 
profile.)

\subsection{New DM profiles}

Adopting a purely phenomenological approach we can readily write down a 
family of DM profiles that satisfy the first two requisite properties, 
resembling an isothermal profile in the inner region and falloff 
asymptotically as $r^{-3}$ at large radii. These profiles can be 
characterized by a scale radius $r_a$ (which is generally different from 
the NFW scale radius $r_s$), and the set of three parameters 
$(\eta,\nu$,$\lambda)$:
\begin{equation}
\rho=\frac{\rho_0^{\star}}{\left[1+x^\eta
  \right]\left[1+x^\nu\right]^\lambda}\,,
\end{equation}
for which $\eta + \nu + \lambda = 3$, and $x=r/r_a$. Here we consider the 
four simplest profiles characterized by ($\eta,\nu$,$\lambda$)= (0,1,3), 
(1,2,1), (0,2,3/2) and (3,$\nu$,0), with $\rho_0^{\star} = 2\rho_0$ for
$\eta = 0$, and $\rho_0^{\star} = \rho_0$ for $\eta = 1, 3$. The second 
profile was previously deduced to provide a good fit to the distribution of 
DM in dwarf galaxies (Burkert, 1995).

\begin{figure}
\begin{center}
\includegraphics*[width=8cm]{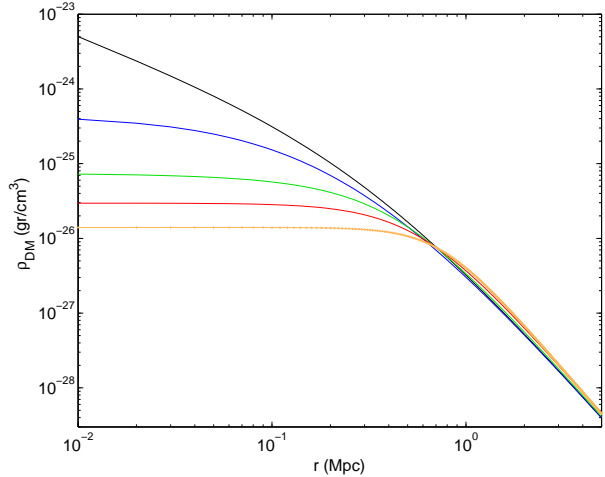}
\caption{The DM profiles $\rho^I-\rho^{IV}$ are
  plotted in blue, green, red and orange lines, respectively; the
  black line shows the NFW profile.}
\end{center}
\end{figure}

In order to be able to meaningfully compare the different models, we 
take a nominal value of $10^{15}$ M$_{\odot}$ for the total mass 
of the cluster (mostly that of the DM and gas) at the virial radius, 
$r_{vir}\approx1.5\;Mpc$. Doing so relates the central density and 
scale radius, so the values of these quantities cannot be 
arbitrarily selected. The requirements of the equality of the total
masses at the virial radius and the $r^{-3}$ falloff at large radius
completely specify the profile parameters. 
The four new DM profiles are shown in figure 4 together with the NFW 
profile. Clearly, values of the central densities span a wide range, 
and the convergence of the first profile ($\rho^I$) to the 
asymptotic $r^{-3}$ law is the fastest.

\begin{figure}
\begin{center}
\includegraphics*[width=8cm]{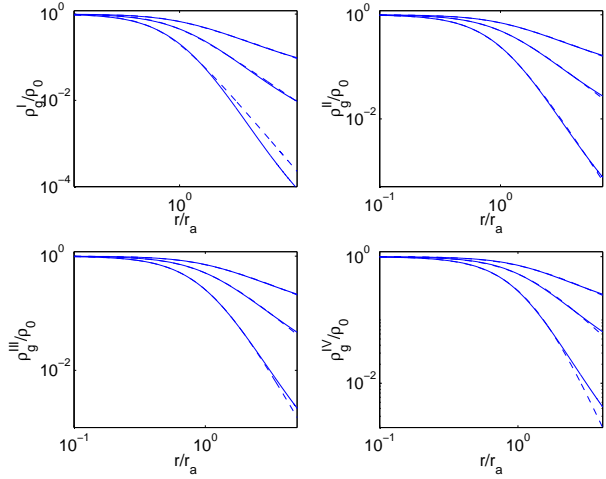}
\caption{Gas profiles obtained from substituting the new DM
  profiles into the HE equation (solid lines) and the best
  fit to a $\beta$-profile (dashed lines). Results for the first 
  and second profiles are shown in the upper right and left 
  panels, and those for the third and fourth profiles are in 
  lower right and left panels, respectively. In each panel the 
  upper to lower graph are for the cases corresponding to 
  $B=5,\;10$, and $20$, respectively.}
\end{center}
\end{figure}

To check whether the third condition is also satisfied, we substitute the 
new profiles into the HE isothermal equation, neglecting the gas and galaxy 
contributions to the total mass, and obtain the following gas profiles,
\begin{equation}
\rho_g^i(x)=\rho^i_{g0}\exp[-B f^i(x)]
\end{equation}
where
\begin{eqnarray}
f^I(x)&=&\frac{2+x}{2+2x}-\frac{\ln(1+x)}{x}\\
f^{II}(x)&=&\frac{1}{4x}\left\{2(1+x)\left[\arctan(x)-\ln(1+x)\right]+(x-1)\ln(1+x^2)\right\}\\
f^{III}(x)&=&1-\frac{arcsinh(x)}{x}\\
f^{IV}(x)&=&\frac{1}{6x}\left[ 3 x^3 \;_2F_1\left(
 \frac{2}{3},1,\frac{5}{3},-x^3\right)-2 \ln(1+x^3)\right]
\end{eqnarray}
where $_2F_1$ is the hypergeometric function. 
\noindent
The parameter B was, defined in eq. (9), depends on the DM 
parameters $r_a$ and $\rho_0$, and the gas parameters $T_0$
and $\beta$. 
Due to the stipulated constancy of the mass at the virial radius, $B$ 
essentially depends only on the gas temperature, $T_0$. In Table 2 
we list values of B for temperatures in the range 5-15 keV.

\begin{table}
\begin{center}
\caption{Values of B characterizing the DM profile for central 
temperatures in the range of 5-15 keV.}
\begin{tabular}{|c|c|c|c|c|} \hline\hline
$B_{NFW}$ & $B_I$ & $B_{II}$ & $B_{III}$ & $B_{IV}$ \\\cline{1-5}
20 & 35.13 & 18.94 & 13.64 & 11.39 \\\cline{1-5}
10 & 17.56 & 9.47 & 6.82 & 5.69 \\\cline{1-5} 
5 & 8.78 & 4.73 & 3.41 & 2.84 \\\cline{1-5}
\end{tabular}
\end{center}
\end{table}     

Fits of the new gas profiles to a $\beta$ model are shown in 
figure 5.
Although all the fits provide better approximation to the $\beta$ 
model than the corresponding fit from the NFW model, values of the 
fit parameters for the third and fourth profiles are somewhat unlikely.  
The second profile yields good results only on small scales 
($r\leq 0.7-0.8\;Mpc$); the fit is poor at larger radii. The
best fit is obtained with the first profile. 

We checked whether each of the above DM profiles can be closely 
approximated by fitting a solution of the HE equation obtained with 
an isothermal $\beta$ profile for the gas. The fit parameters were 
$\tilde{A}, r_c, \rho_{g0}$ and $\beta$. Not surprisingly, we find that 
all of the above four profiles can be very well fit by a solution of 
the HE equation. Here again the fit to the NFW profile is very poor. 
We have also compared the gas profiles deduced from the HE equation for 
each of the DM models directly to the observed quantity by numerically 
evaluating the SB. (In these computations the gas density was truncated 
at $x=20$, corresponding to a limiting radius which is larger than the 
virial radius.) 

Results of the fits of the deduced SB profiles to a $\beta$ model 
are shown in Figure 6. The first two profiles yield good fits to 
a SB that has a form of a $\beta$ profile, with reasonable parameter 
values. The third and fourth profiles are better fit by a 
function of the form of $S_x^{new}$ (Eq. 10), with $\xi\neq 2$ and 
$\tilde{\beta}\neq 3 \beta-1/2$, as does the modified NFW model.

Next we derived the gas density distributions from the HE 
equation for polytropic gas and with the DM mass corresponding to 
the above profiles. The deduced gas density profiles are 
represented in terms of the function $f(r)$ in eq. (11). Since 
this function is monotonically increasing with $r$, there is 
an upper limit on the value of $\gamma$ below which the density 
is positive definite. We have determined the limiting values of 
$\gamma$ for typical values of B and $r_{max}=1.5 \,Mpc$. 
Only for the first DM profile these limits are acceptable. For 
the other three profiles the deduced values of $\gamma$ are 
either unrealistic or negative, and thus unacceptable for the 
observed range of gas temperatures.

As discussed previously, the mass obtained from the HE equation 
for isothermal gas closely approximates the cluster total mass 
since the gas is approximately isothermal and the fractional mass 
contribution of the gas is small. We have compared the DM masses 
obtained from direct integrations of the DM density distributions 
and those obtained from the HE equation for isothermal gas. 
For isothermal gas, the deduced cluster mass is
\begin{equation}
M^{iso}(r)=\frac{3 k T_0 \beta}{\mu m_p G r_c^2}\frac{r^3}{\left(1 
+\frac{r^2}{r_c^2}\right)}\,.
\end{equation}

\begin{figure}
\begin{center}
\includegraphics*[width=6cm]{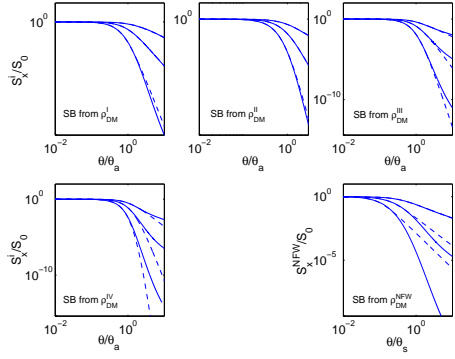}
\caption{The SB obtained from substituting each DM
  profile into $S_x$ is shown by a solid line, and the best fit
  to a SB $\beta$-profile is plotted as a dashed line. The
  cases corresponding to $B=10,\;15$ and $20$ are shown in each 
  figure from the upper to the lower lines, respectively. Also 
  plotted are the results for the SB corresponding to the NFW 
  profile.}
\end{center}
\end{figure}

The DM mass profiles for the new profiles can written as
\begin{equation}
M(r)=4 \pi \rho_0 r_a^3 m(x)\,,
\end{equation}
where $m(x)$ are obtained by integrating the four profiles:
\begin{eqnarray}
m^I(x)&=&\ln\left(1+x\right)-\frac{x\left(2+3x\right)}{2\left(
1+x\right)^2}\\
m^{II}(x)&=&\frac{1}{4}\left[2\ln(1+x)+\ln(1+x^2)-2\arctan(x)
\right]\\
m^{III}(x)&=&arcsinh(x)-\frac{x}{\sqrt{1+x^2}}\\
m^{IV}(x)&=&\frac{1}{3}\ln(1+x^3) .
\end{eqnarray}
The new profiles can be well described by such a function, while only 
a poor fit is obtained to the NFW mass profile.

Finally, the presence of gravitational arcs in some clusters implies
sufficiently high central mass densities. We have evaluated the
central surface densities for the new profiles using the 
Abel integral; these are
\begin{eqnarray}
\Sigma^I(0)&=&\rho_0^I r_a^I \,\\
\Sigma^{II}(0)&=&\frac{\pi}{2} \rho_0^{II} r_a^{II} \,\\
\Sigma^{III}(0)&=&2\rho_0^{III} r_a^{III}\,\\
\Sigma^{IV}(0)&=&\frac{4 \pi}{3 \sqrt{3}} \rho_0^{IV} r_a^{IV}
\end{eqnarray}
for the first to the fourth profile. Actual values are 
$\sim 1$ g/cm$^{2}$ for the first DM profile, to about a factor 
$\sim 4$ lower for the fourth profile. Since these values are quite 
comparable to the critical density $\Sigma_c \approx 0.5$ g/cm$^{2}$ 
needed to produce arcs, no additional constraint is imposed on these 
models by this consideration.

\begin{table*}[t]  
\caption{Parameters of DM profiles from fits to the SB data, and 
  reduced $\chi^2$ of the fits. $\sigma_{\chi}$ is the standard
  deviation for the reduced $\chi^2$,
and the scale radius is in Mpc.}
\vspace{0.2cm}
\small
\begin{tabular}{ l@{\hspace{0.5cm}} c@{\hspace{0.5cm}} c@{\hspace{0.5cm}} 
c@{\hspace{0.5cm}} c@{\hspace{0.5cm}} c@{\hspace{0.5cm}}
c@{\hspace{.5cm}} c@{\hspace{.5cm}} c@{\hspace{.5cm}} } \hline
cluster & z & \multicolumn{3}{c}{Fitting Parameters}  &
\multicolumn{4}{c}{Goodness of the fit} \\
 & & $r_a, \, B_I$ & $r_a, \, B_{II}$ & $r_s, \, B_{N}$ & $\chi_{I}^2$ &
 $\chi_{II}^2$ & $\chi_{N}^2$ & $\sigma_{\chi}$  \\\cline{1-9}
A401    &       0.0748  &       0.29  ,  14.88   &       0.32  ,  8.51    &       0.87  ,  8.92    &       1.69    &       2.33    &       1.92    &       0.20\\
A478    &       0.0881  &       0.17  ,  16.28   &       0.19  ,  9.35    &       0.50  ,  9.61    &       5.71    &       18.75   &       2.98    &       0.35\\
A520    &       0.203   &       0.59  ,  19.31   &       0.56  ,  10.08   &       2.73  ,  15.01   &       1.19    &       1.03    &       1.70     &       0.33\\
A545    &       0.153   &       0.45  ,  19.31   &       0.44  ,  10.39   &       1.66  ,  13.1    &       1.49    &       0.97    &       2.89    &       0.45\\
A586    &       0.171   &       0.22  ,  16.43   &       0.24  ,  9.27    &       0.68  ,  9.88    &       1.38    &       1.10     &       1.98    &       0.33\\
A644    &       0.0704  &       0.24  ,  16.07   &       0.25  ,  9.10     &       0.68  ,  9.50     &       2.87    &       2.45    &       4.30     &       0.28\\
A1413   &       0.1427  &       0.22  ,  16.24   &       0.24  ,  9.21    &       0.69  ,  9.86    &       1.97    &       2.94    &       2.10     &       0.32\\
A1651   &       0.0825  &       0.24  ,  15.67   &       0.25  ,  8.82    &       0.63  ,  8.96    &       1.82    &       1.68    &       1.72    &       0.28\\
A1656 & 0.0232 & 0.54 , 18.76 & 0.54 , 10.29 & 2.20 , 13.09 & 1.16 & 0.98 & 3.39 & 0.32\\
A1689   &       0.181   &       0.22  ,  17.55   &       0.24  ,  9.94    &       0.70  ,   10.71   &       2.69    &       5.40     &       2.08    &       0.3\\
A1763   &       0.187   &       0.36  ,  15.79   &       0.38  ,  8.90     &       1.15  ,  9.76    &       4.03    &       6.32    &       2.36    &       0.39\\
A2029   &       0.0765  &       0.14  ,  15.06   &       0.16  ,  8.71    &       0.40   ,  8.80     &       3.55    &       11.15   &       4.19    &       0.39\\
A2163   &       0.203   &       0.41  ,  15.86   &       0.42  ,  8.75    &       1.40   ,  10.21   &       1.37    &       1.35    &       2.52    &       0.34\\
A2204   &       0.1523  &       0.15  ,  15.96   &       0.17  ,  9.22    &       0.43  ,  9.29    &       0.83    &       1.29    &       0.93    &       0.38\\
A2218   &       0.175   &       0.34  ,  17.21   &       0.35  ,  9.49    &       1.19  ,  11.17   &       1.00       &       1.30     &       2.76    &       0.31\\
A2244   &       0.097   &       0.18  ,  15.68   &       0.20  ,   9.04    &       0.54  ,  9.30     &       1.06    &       1.23    &       1.44    &       0.24\\
A2255   &       0.0809  &       0.86  ,  21.53   &       0.76  ,  10.62   &       6.91  ,  26.09   &       0.99    &       1.14    &       0.95    &       0.21\\
A2319   &       0.0559  &       0.31  ,  14.12   &       0.36  ,  8.15    &       0.94  ,  8.49    &       3.02    &       5.21    &       1.46    &       0.22\\
A2507   &       0.196   &       0.58  ,  16.04   &       0.56  ,  8.52    &       2.97  ,  13.39   &       1.02    &       0.95    &       1.08    &       0.21\\
A3112   &       0.0746  &       0.11  ,  15.23   &       0.15  ,  9.02    &       0.29  ,  8.85    &       1.44    &       2.15    &       1.61    &       0.27\\
A3667   &       0.0542  &       0.32  ,  13.29   &       0.34  ,  7.49    &       1.02  ,  8.22    &       4.99    &       7.81    &       5.33    &       0.35\\
A3888   &       0.168   &       0.47  ,  22.09   &       0.47  ,  12.24   &       2.38  ,  18.15   &       1.54    &       1.65    &       1.60     &       0.16\\
PKS0745 &       0.1028  &       0.13  ,  16.19   &       0.15  ,  9.42    &       0.38  ,  9.52    &       2.85    &       3.47    &       4.65    &       0.25\\
Triang  &       0.051   &       0.38  ,  15.87   &       0.41  ,  8.91
&       1.85  ,  9.84    &       3.86    &       6.29    &       2.06
&       0.23\\
\cline{1-9}
\end{tabular}
\end{table*}

\section{Parameters of the DM profiles from a ROSAT Sample}

In an attempt to further distinguish between the four DM profiles 
described in the previous section and possibly select the most 
realistic profile based on available observational data, we have 
used results from a sample of SB profiles of clusters measured 
with the ROSAT PSPC. The sample  -- a subset of a dataset which was 
compiled and investigated by Ettori \& Fabian (1999; the data were 
kindly provided by Ettori) -- consists of 24 clusters with X-ray 
luminosities $\geq 10^{45}$ erg $s^{-1}$ (taking $H_0=50\;
km\;s^{-1}\;Mpc^{-1}$), at redshifts in the range $0.051-0.203$. 
We have selected nearby and moderately distant  
clusters for which the ROSAT PSPC provides some -- albeit not 
optimal -- spatial resolution (for more details, see Ettori \& 
Fabian 1999). 
We performed fits of the measured SB profiles to those predicted 
from three different models for the isothermal gas 
density. The assumption of gas isothermality is a reasonable 
approximation to the temperature profile at radii larger than 
$~0.1r_{vir} \; (\approx 0.2Mpc)$. All three models were deduced 
from the HE equation adopting these DM profiles: NFW, and our 
first ($\rho^I$) and second ($\rho^{II}$) models. The corresponding 
gas density distributions are
\begin{equation}
\rho_g^{N}(x)=\rho_{g0}^{N}\exp^{-B_{N}}(1+x)^{B_{N}/x}
\,,
\end{equation}
where $x=r/r_s$,
\begin{equation}
\rho_g^I(x)=\rho_{g0}^I\exp^{-\frac{B_I(2+x)}{2+2x}}(1+x)^{B_I/x} 
\,,
\end{equation}
and 
\begin{equation}
\rho_g^{II}(x)=\rho_{g0}^{II}\exp^{-\frac{B_{II}}{4x}\left\{2(1+x)\left[\arctan(x)-\ln(1+x)\right]+(x-1)\ln(1+x^2)\right\}} \,,
\end{equation}
where $x=r/r_a$. 
The fits were performed by $\chi^2$ minimization (using the 
'Minuit' CERN program).

Results of the fits are summarized in Table 3, where in addition 
to listing the best-fit values of the scale radii and $B$, values 
of the reduced $\chi^2$ ($\chi^2/dof$) and the standard deviation for
the reduced $\chi^2$ are also specified. Of the three models, the
first DM profile provides the best fit to the data of 11 out of the 
24 clusters, with the second profile providing the best fit in 8 
clusters. Furthermore, for most of the latter 8 clusters the 
differences between the quality of the fits based on the first and 
second DM models are not statistically significant. Based on the 
results from this ROSAT dataset it is apparent that the first 
DM profile is most consistent with the data while the NFW profile 
is the least favored. (We note that all three fits to the data 
on three clusters -- A478, A1763 \& A3667 -- are very poor, 
raising doubts on the validity of the assumption of hydrostatic 
equilibrium of the gas in these clusters, perhaps due to ongoing 
merger activity?).

In order to assess the impact of a more precise but similar 
database, we have repeated the fits by artificially reducing the 
observational errors in the measurements of the SB. Doing so 
does not affect appreciably the quality of the fit to the SB 
from the first DM model, but reduces the consistency with the 
second DM model and significantly worsening the viability of 
the NFW profile. Clearly, since this test is based on the 
current database it does not add independent confirmation 
of the results, but rather just a simulation of 
what might be feasible to do when higher quality data become 
available.

\section{Discussion}

The aim of this work has been to find alternative DM profiles that 
are finite at the cluster center and are consistent with 
observed X-ray SB profiles. 
Our approach is purely 
phenomenological and is based on the selection of simple 
non-cusped profiles that falloff asymptotically as $\propto r^{-3}$ 
at large $r$. We first constructed two modified NFW profiles 
($\rho^{iso}$ and $\rho^{poly}$) by truncating the NFW below 
some inner radius ($r_b$) merely to remedy the central divergence 
of the NFW model. Since $r_b \sim 50$ kpc, the new profiles 
quickly converge to the NFW profile. But the requirement that 
the related gas density profile has an associated thermal 
bremsstrahlung SB with the typical $\beta$ model shape led us 
to abandon these modified profiles as realistic alternatives 
to the NFW model.

We then considered four new profiles characterized by the 
three parameters ($\mu,\nu,\lambda$) with the requisite features, 
and tested their viability by contrasting their associated 
SB profiles with ROSAT data on a sample of 24 clusters. 
Comparison of the gas profiles resulting from the new DM 
models with $\beta$ gas profiles 
clearly shows that all of these give better results than the NFW 
profile. Below a radius of about 1 Mpc, the behavior of all four 
profiles is acceptable, but progressively degrades at larger radii. 
Examining the SB profile calculated from the different DM 
profiles we saw that the results are not unequivocal, namely 
that the general shape of the SB function can be fitted quite 
well to a $\beta$ SB profile. This is so for a fit done over a 
large range of radii; the fit is particularly good over the 
radial range $r\leq 1.5-2\;Mpc$. The reason for this is the fact 
that a SB that has the shape of a $\beta$ profile necessarily has a
flat slope in the central region. Upon detailed comparison 
of the results of the fits, as well as consistency with polytropic 
gas distributions,
we concluded that the first of these four profiles, for which 
$(\eta,\nu,\lambda)=(0,1,3)$: 
\begin{equation}
\rho(r)=\frac{\rho_0}{\left(1+r/r_a\right)^3}\,,
\end{equation}
is most consistent with the ROSAT sample. We emphasize, however, 
that due to the large observational uncertainties the preference 
of the first profile over the second is not 
large.
It is quite likely that the availability of more precise 
X-ray SB and temperature measurements 
will enable a more definite distinction to be made between 
these viable alternatives to the NFW profile.

Independent recent work (El-Zant, Shlosman \& Hoffman 2001) also 
leads to a resolution of the 'core catastrophe' in galaxies -- 
the discrepancy between the diverging inner density profile of 
DM from CDM N-body simulations and the finite core deduced from 
observations. These authors suggest the gas in galaxies is not 
initially smoothly distributed in the DM halo, but rather is 
concentrated in small clumps containing $\sim 0.01\%$ of the 
total mass. The orbital energy of the clumps dissipates by  
dynamical friction as they move in background of DM particles, 
thereby transferring energy to the DM and heating it. 
This process is said to be sufficiently effective to turn the 
primordial cusp of the DM profile into a non-diverging core, 
resulting in a profile of the form 
\begin{equation}
\rho=\frac{C}{\left(r+r_c\right)\left(A+r\right)^2}\,,
\end{equation}
where $C$ is a fixed parameter, $A$ and $r_c$ are a scale parameter
and core radius, respectively. Furthermore, it was found that best
fits require that $A=r_c$, which then yields essentially the same 
profile that we have deduced for clusters. This similarity between 
the behavior of the DM profile in galaxies and in clusters is 
indeed expected in theories of formation and evolution of the 
large scale structure. Gravitational drag could also be important in
clusters (e.g., Rephaeli \& Salpeter 1980), so a similar process of
transfer of kinetic energy of the galaxies (initially with their DM halos)
to the IC DM could have flattened also cluster profiles.

Flattening of DM profiles in the centers of galaxies and clusters 
is suggested by other theoretical considerations. For example,
D'Onghia, Firmani \& Chincarini (2002) have recently argued that the 
flattening can occur in galactic and cluster centers if the DM in
these systems consists of weakly self-interacting particles. Collisions 
between the particles during system collapse and the associated inward
transfer of heat lead to expansion of the core. They propose that this
process is implemented in N-body simulations by modifying the initial 
conditions and taking a self-interaction cross section that is
inversely proportional to the particle velocity.

Finally, the physical motivation to find a more acceptable form for DM 
density profiles in galaxies and clusters, and the already available 
observational data, provide a viable basis for selecting between simple, 
well-behaved profiles. We have identified what seems to be the most 
consistent form of the DM distribution in clusters. Our work has been 
based on a simplified theoretical description of clusters -- such as 
the sphericity of the cluster and isothermality of IC gas -- assumptions 
that can be relaxed when higher quality spectral and spatial XMM 
and {\it Chandra} measurements of clusters will be available.

\section*{ACKNOWLEDGEMENTS} 
We are grateful to Dr. Stefano Ettori for providing the ROSAT dataset 
discussed in this paper.

\end{document}